\documentclass[amsmath,amssymb,nofootinbib,prl]{revtex4}

\usepackage{setspace}
\usepackage{graphicx}

\usepackage{amsmath,amssymb,amsfonts,amsthm,mathrsfs}
\usepackage{enumerate}

\begin{document}

\title{Towards a quantum notion of covariance in spherically symmetric loop quantum gravity}

\author{Rodolfo Gambini$^1$, Javier Olmedo$^2$ Jorge Pullin$^3$}
\affiliation{1. Instituto de F\'{\i}sica, Facultad de Ciencias, Igu\'a 4225, esq. Mataojo,
11400 Montevideo, Uruguay. \\
2. Departamento de F\'{\i}sica te\'orica y del cosmos, 
Universidad de Granada, Granada-18071, Spain\\
3. Department of Physics and Astronomy, Louisiana State University,
Baton Rouge, LA 70803-4001, USA.}

\begin{abstract}
The covariance of loop quantum gravity studies of spherically symmetric space-times has recently been questioned. This is a reasonable worry, given that they are formulated in terms of slicing-dependent variables. We show explicitly that the resulting space-times, obtained from Dirac observables of the quantum theory, are covariant in the usual sense of the way ---they preserve the quantum line element--- for any gauge that is stationary (in the exterior, if there is a horizon). The construction depends crucially on the details of the Abelianized quantization considered, the satisfaction of the quantum constraints and the recovery of standard general relativity in the classical limit  and suggests that more informal polymerization constructions of possible semi-classical approximations to the theory can indeed have covariance problems. This analysis is based on the understanding of how  slicing dependent quantities as the metric arise in a quantum context in terms of parameterized observables. It has implications beyond loop quantum gravity that hold for general approaches to quantum space time theories.

\end{abstract}
\maketitle

The application of loop quantum gravity techniques to spherically symmetric space-times  has led to insights about how the singularity inside black holes could be eliminated by quantum effects. We refer specifically to the construction that uses inhomogeneous slices and enforces the constraint algebra at the quantum level through Abelianization \cite{us}. The construction is based on canonical quantum gravity and, as such, is based on three dimensional objects that are slicing-dependent. A reasonable worry \cite{bojo} is if the constructions lead to covariant quantizations of the space-times. An encouraging sign is that they enforce the constraint algebra, which in the canonical theory is the guarantor of slicing independence of the construction, and of reproducing the standard general relativistic results in the classical limit. However, technical aspects, like the fact that the algebra is Abelianized, may lead to questions about the covariance of the procedure. We would like to show explicitly that the resulting space-times are indeed covariant at any space time region in the usual sense of the word: the invariant line element is indeed invariant at the quantum level for any stationary foliation (stationary in the exterior if there is a horizon). Although this is not a definitive extension of the notion of  covariance to the quantum realm, the result arises in a non-trivial way and suggests that with more work this notion could be achieved.

We start with a spherically symmetric space-time. Following the discussion in \cite{us}, the line element can be written as $dS^2=ds^2+|E^x| d\omega^2$, where $d\omega^2$ is the line element of the unit 2-sphere, and,
\begin{equation}
ds^2=-(N^2-N_xN^x)dt^2+2N_xdtdx+\frac{(E^\varphi)^2}{|E^x|}dx^2,
\end{equation}
where $N$ and $N_x$ are suitable lapse and shift functions, with $N^x=g^{xx}N_x$, and $E^\varphi$ and $E^x$ triad variables, conjugate to the extrinsic curvature components of the foliation, $K_\varphi$ and $K_x$, with Poisson brackets,
\begin{align}\nonumber\label{eq:poiss}
&\{K_x(x),E^x(x')\}=G\delta(x-x'),\\
&\{{K}_\varphi(x),E^\varphi(x')\}=G\delta(x-x').
\end{align} 
We take the Immirzi parameter $\gamma=1$ and $G$ is Newton's constant.

We re-define the lapse and the shift in order to make the constraints Abelian as shown in \cite{us-improv}
\begin{eqnarray}
\overline{N}^x&=&N^x-\frac{2 N K_\varphi\sqrt{\vert E^x\vert}}{\vert E^x\vert'},\\
\overline{N}&=& -\frac{1}{E^\varphi}\left(N \frac{E^\varphi}{\vert E^x\vert'}\right)',
\end{eqnarray}
(prime denotes derivative with respect to $x$ with which the smeared Hamiltonian constraint takes the form,
 \begin{equation}\label{eq:H_new-den}
\tilde H(\tilde N) :=\frac{1}{G}\int dx \tilde N
 \sqrt{\vert E^x\vert}E^\varphi\bigg[ K_\varphi^2-\frac{[\vert E^x\vert']^2}{4
 	(E^\varphi)^2}+\left(1-\frac{2 G M}{\sqrt{\vert E^x\vert}}\right)\bigg],
 \end{equation}
where $M$ is the ADM mass.\footnote{Classically, the variational problem is well-posed once a boundary term is introduced \cite{kuchar,bh-review}. The boundary term explicitly introduces the mass $M$ and its conjugate variable $\tau$ (the proper time of an asymptotic observer) as the global physical degree of freedom that characterizes classical solutions.} It can be checked that this constraint has an Abelian algebra with itself. The momentum constraint, on the other hand, keeps its original form.

Let us recall the construction of the physical Hilbert space. The elements of a basis of quantum states are one dimensional spin networks with integer valences $k_j$ at each link $j\in[-S,-S+1,\ldots,0,\ldots,S-1,S]$, with $2S+1$ nodes in total. We have three kinds of Dirac observables, one corresponding to the ADM mass $M$, the total number of vertices $S$ and the set of integers $\vec{ k}$,
\begin{equation}
    \hat{M}\vert M,\vec{k}\rangle = M\vert M, \vec{k}\rangle,\label{6}
\end{equation}
and the other two not having a classical counterpart,
\begin{equation}
    \hat{O}(z) \vert M,\vec{k}\rangle = \ell^2_{\rm Planck} k_{{\rm Int}(S z)} \vert M,\vec{k}\rangle
\end{equation}
where ${\rm Int}$ means the integer part and $z$ is a real parameter in the interval $[-1,1]$, so $O(z)$ constitutes a one-parameter family of observables. $M$, $O(z)$ and $S$ are the Dirac observables. $\ell_{\rm Planck}$ is Planck's length. Having identified the physical space, we will describe the metric by introducing gauge fixings that allow to write the metric components in specific gauges in terms of evolving observables defined on the physical space of states.

The action of the triads and their derivatives on physical states is,
\begin{eqnarray}
\hat{\vert E^x}(x_j)\vert\vert M, \vec{k}\rangle &=& \hat{O}\left(z\left(x_j\right)\right) \vert M,\vec{k}\rangle=
\ell_{\rm Planck}^2 k_j \vert M, \vec{k}\rangle,\\
\vert\hat{E}^x(x_j)\vert' \vert M, \vec{k}\rangle &=& \frac{1}{\ell_{\rm Planck}}(\vert \hat{E}^x\vert(x_{j+1}) -\vert\hat{E}^x\vert(x_j)) \vert M, \vec{k}\rangle. 
\end{eqnarray}
Here, and for simplicity, we choose a particular class of spin networks, with given $M$ and $k_j$ (no superpositions) and  gauge fixing such that $x_j^2 =\ell_{\rm Planck}^2 k_j$, where
\begin{equation}
x_j=\ell_{\rm Planck}\left(|j|+j_0\right).   \label{8}
\end{equation}
This is motivated in that in the classical theory the condition $\vert E^x\vert=x^2$ corresponds to having $x$ be the radius of the spheres of symmetry. This includes many popular coordinate systems for studying spherical space-times. We will later relax this assumption, allowing arbitrary stationary changes in the radial coordinate.

Within the improved dynamics of \cite{us-improv}, $j_0$ is the minimum integer greater than $\left(\frac{2GM\Delta}{4 \pi l^3_{\rm Planck}}\right)^{1/3}$ and $\Delta$ is the loop quantum gravity area gap \cite{us-improv}. This implies,
\begin{eqnarray}
\vert\hat{E}^x(x_j)\vert\vert M, \vec{k}\rangle &=& x_j^2 \vert M\,\vec{k}\rangle,\\
\vert\hat{E}^x(x_j)\vert' \vert M, \vec{k}\rangle &=& {\rm sign}(j)\left(2 x_j +\ell_{\rm Planck}\right)\vert M, \vec{k}\rangle,
\end{eqnarray}
where we take ${\rm sign}(j)=1,\,\forall j\ge0$ and ${\rm sign}(j)=-1,\,\forall j<0$. 

%To quantize the constraints we proceed to polymerize,
The quantization of the constraints involves,
\begin{equation}
    K^\varphi \longrightarrow \frac{\sin\left(\rho_j K_\varphi(x_j)\right)}{\rho_j},
\end{equation}
with $\hat \rho_j^2 =\Delta/(4\pi \hat E^x(x_j))$.  The quantum gauge fixings that we adopt here leave  $K_\varphi(x_j)$ as  either a c-number function or a function of the  Dirac observables $M$, $S$ or $O(z)$. These quantum gauge fixings correspond to a choice of a slicing in the quantum theory. 

%Generically, 
Moreover, the conjugate variable to $K_\varphi(x_j)$ on the physical Hilbert space is obtained by solving the Abelianized constraint
(\ref{eq:H_new-den}), and amounts to,
%, we have that,
\begin{equation}
\left[\hat E^\varphi\left(x_j\right)\right]^2 =
{\left[\vert\hat E^x\vert(x_j)'\right]^2}
\left[4\left(1 +
\frac{\widehat{\sin^2\left(\rho_j K_\varphi(x_j)\right)}}{\hat \rho_j}
-\frac{2 G \hat M}{\sqrt{\vert\hat E^x\vert(x_j)}}
\right)\right]^{-1}.
\end{equation}
From now on we will use the notation $\hat E^x_j\equiv  \vert \hat{E}^x\vert(x_j)$, $(\hat E^x_j)'\equiv  \vert\hat{E}^x\vert(x_j)'$, $\hat E^\varphi_j\equiv \hat E^\varphi(x_j)$ and $K_{\varphi,j}\equiv K_\varphi(x_j)$.  The metric of space-time can be implemented as a quantum operator acting on the physical space of states by writing it as a parameterized observable  given by the gauge fixing conditions. Details are in \cite{us}.

Let us consider stationary slices, that is, the gauge fixing conditions do not depend on time. Preservation of these conditions \cite{us-improv} correspond to $\overline{N}^x=\overline{N}=0$ and imply that $N E^\varphi/( E^x)'$ is a constant that we take equal to $1/2$. This in turns means that $N=(E^x)'/(2E^\varphi)$ and this implies that $N^x=K^\varphi \sqrt{E^x}/E^\varphi$. The gauge fixing determines the Lagrange multipliers that were mentioned before.
    
We will first analyze what happens outside the ``bounce'' that replaces the classical singularity and later study the covariance at the bounce. For the physical states $|M,{\hat k}\rangle$ under consideration (with no superpositions in $M$ and $k_j$), the Schwarzschild metric can be readily obtained by fixing $K_{\varphi,j}=0$. It is given by,\footnote{Note that the line element of the 2-spheres is determined by $g_{\theta\theta}= x_j^2$ and $g_{\varphi\varphi}= x_j^2 \sin^2\theta$.} 
\begin{eqnarray}\label{eq:schld-gtt}
\hat{g}^S_{tt}(x_j)&=& -\frac{\left[\left(\hat{E}^x_j\right)'\right]^2}{4 \left({E}^\varphi_j\right)^2}= -\hat{N}^2 = -\left(1-\frac{\hat{r}_S}{\sqrt{\hat{E}^x_j}}\right),\\
\hat{g}^S_{tt}(x_j) \vert M, \vec{k}\rangle
&=& -\left(1-\frac{\hat{r}_S}{x_j}\right)\vert M, \vec{k}\rangle,\\\label{eq:schld-gtx}
\hat{g}^S_{tx}(x_j)&=&0, \,\,\textrm{since}\,\, K_{\varphi,j}=0,\\\label{eq:schld-gxx}
\hat{g}^S_{xx}(x_j)&=&\frac{\left(E^\varphi_j\right)^2}{\hat{E}^x_j}=
\frac{\left[ \left(\hat{E}^x_j\right)'\right]^2}{4 \hat{E}^x_j} \frac{1}{1-\frac{\hat{r}_S}{\sqrt{\hat{E}^x_j}}} \\
\hat{g}^S_{xx}(x_j)\vert M, \vec{k}\rangle
&=&
\frac{\left(2 x_j+\ell_{\rm Planck}\right)^2}{4 x_j^2}\frac{1}{1-\frac{{r}_S}{x_j}}\vert M, \vec{k}\rangle\nonumber\\
&=& \left(1+\frac{\ell_{\rm Planck}}{x_j}+\frac{\ell_{\rm Planck}^2}{4x_j}\right)
\left(1-\frac{{r}_S}{x_j}\right)^{-1}\vert M, \vec{k}\rangle\nonumber\\
&=&{\left(1+\frac{\ell_{\rm Planck}}{2 x_j}\right)^2}
\left({1 -\frac{{r}_S}{x_j}}\right)^{-1}\vert M, \vec{k}\rangle,
\end{eqnarray}
with $\hat{r}_S=2G\hat{M}$. It should be noted that the calculation is exact. Let us proceed to compare the result with the action of the metric with a generic choice of functional parameter of the  observable $K_{\varphi,j}$ only restricting to stationary foliations (independent on time) that will include usual ones like the Painlev\'e--Gullstrand and Eddington--Finkelstein ones.

On generic stationary foliation, lapse, shift and $E^\varphi_j$ can be written as,
\begin{eqnarray}
  N_j&=&\frac{1}{2}\frac{\left(E^x_j\right)'}{E^\varphi_j},\qquad N^x_j=\frac{\sin\left(\rho_j K_{\varphi, j}\right)}{\rho_j}\frac{\sqrt{E^x_j}}{E^\varphi_j},\\
  E^\varphi_j&=& \left(E^x_j\right)'\left(2\sqrt{1+\frac{\sin^2\left(\rho_j K_{\varphi, j}\right)}{\rho_j^2} -\frac{2 GM}{\sqrt{E^x_j}}}\right)^{-1}
\end{eqnarray}
with the parameter of the observable $K_{\varphi,j}$ generic but time independent. These are the quantum versions of the classical expressions discussed above.

The choice of $K_{\varphi,j}$ completes the prescription for the gauge fixing that characterizes the foliation. We recall that it may be considered as the functional parameter of the parameterized  observable that defines $E_j^\varphi$, and through it the metric components, and therefore can be chosen at will. Each choice determines a different system of coordinates.
One way of doing this is to introduce a function $F(x_j)$ such that $\sin\left(\rho_j K_{\varphi,j}\right)=F(x_j)$ with $F(x_j) \in [-1,1]\,\, \forall x_j$ and therefore, with the notation $F(x_j)\equiv F_j \in [-1,1]$.  Each choice of $F_j$ leads to a different foliation, for instance $F(x_j)=\rho_j\sqrt{r_S/\sqrt{E^x_j}}$ leads to ingoing Painlev\'e--Gullstrand form of the metric \cite{extension} and $F(x_j)=\rho_j{r_S}/\sqrt{E^x_j(1+r_S/\sqrt{E^x_j})}$ to ingoing Eddington-Finkelstein coordinates \cite{us-improv}.\footnote{Outgoing coordinates in both cases are defined similarly with a minus sign in the right hand side of these choices of $F(x_j)$.} Note that as we explained for $K_{\varphi,j}$, $F(x_j)$ it can either be a c-number function or an operator, function of the Dirac observables, and should be treated accordingly.

For a generic stationary foliation given by $F(x_j)$
we have,
\begin{eqnarray}
  \hat{N}_F(x_j)&=& \sqrt{1+\frac{F_j^2}{\rho_j^2} -\frac{r_S}{\sqrt{\hat{E}^x_j}}},\\
  \hat{N}^x_F(x_j)&=& \frac{2F_j}{\rho_j} \frac{\sqrt{\hat{E}^x_j}}{\left(\hat{E}^x_j\right)'} \hat{N}_F(x_j),\\\label{eq:Fgxx}
  \hat{g}^F_{xx}(x_j)&=& \frac{\left(\left(\hat{E}^x_j\right)'\right)^2}{4\hat{E}^x_j} \hat{N}_F(x_j)^{-2},\\\label{eq:Fgtt}
  \hat{g}^F_{tt}(x_j)&=&-1+\frac{\hat{r}_S}{\sqrt{\hat{E}^x_j}},\\\label{eq:Fgtx}
  \hat{g}^F_{tx}(x_j)&=& \hat{g}_{xx} \hat{N}^x_F(x_j).
\end{eqnarray}

We would like to show that the length of a space-time curve $\left(t(x),x\right)$ is invariant. 
If the state of the black hole system is given by the basis element $|M,{\hat k}\rangle$ defined in Eqs. (\ref{6}-\ref{8}), to each function $t(x)$ corresponds, in Schwarzchild coordinates, a polygonal curve in the plane $(t,x)$ described by a discrete set of points $[...(t_j,x_j),(t_{j+1},x_{j+1})...]$ where $\sqrt(\hat{E}^x_j)|M,{\hat k}\rangle=\ell_{\rm Planck}\sqrt{k_j}|M,{\hat k}\rangle=(|j|+j_0) \ell_{\rm Planck}|M,{\hat k}\rangle=x_j|M,{\hat k}\rangle$ and
${\widehat t(x_j)}|M,{\hat k}\rangle=t(x_j)|M,{\hat k}\rangle$. More general polygonal curves may be defined by composition of these curves. To be able to discuss changes of slicings in a situation where space is discrete it is necessary to consider polygonal curves in space-time

We assume that the quantum version of the invariant interval between two successive points of the polygonal curve acting on a basis element $|M,{\vec k}\rangle$ of the physical space of states may be written as
\begin{equation}{(\widehat{\Delta s_j}})^2={\widehat{g_{ab}(t_j,x_j)}}\widehat{\Delta {x^a_j}} \widehat{\Delta {x^b_j}},
\end{equation}
with ${\widehat{\Delta x^0_j}}={\hat t}_{j+1}-{\hat t}_j={\widehat{ \Delta t}_j}$ and ${\widehat{ \Delta x^1_j}}={\hat x}_{j+1}-{\hat x}_{j}={\widehat{ \Delta x}_j}$.

In particular, the invariant interval between two successive points of the polygonal in Schwarzschild coordinates $\Delta s^S_j$  and generic stationary coordinates $\Delta s^F_j$ is,

\begin{eqnarray}
  (\Delta s^S_j)^2&=&-\left(1-\frac{r_S}{\sqrt{E^x_j}}\right)\Delta t^2_j+\frac{\left(\left(E^x_j\right)'\right)^2}{4 E^x_j}\frac{1}{1-\frac{r_S}{\sqrt{E^x_j}}} \Delta{x_j}^2 
  ,\label{S}\\
  (\Delta s^F_j)^2&=&-\left(1-\frac{r_S}{\sqrt{E^x_j}}\right)\Delta t_{F,j}^2+ 2 \frac{F_j}{\rho_j} \frac{\left(E^x_j\right)'}{2 \sqrt{E^x_j}} \frac{1}{N_F(x_j)} \Delta t_{F,j} \Delta{x_j}\nonumber\\ &&+\frac{\left(\left(E^x_j\right)'\right)^2}{4 E^x_j} \frac{1}{N_F(x_j)^2} \Delta{x_j}^2,\label{29}
\end{eqnarray}
where we have omitted the hats indicating that $E^x_j$, $N_j$, $\Delta s_j^{S,F}$, $r_S$ and $F_j$ are operators. By now the reader should notice by context what we are referring to with the expressions.
So setting ${\hat t_F(x_j)}=\hat t(x_j)-{\hat a(x_j)}$ and taking into account [\ref{S}],
\begin{eqnarray}
  (\Delta s^F_j)^2 &=& -\left(1-\frac{r_S}{\sqrt{E^x_j}}\right) \Delta t_{F,j}^2+ 2\frac{\Delta a_j}{\Delta x_j} \left(1-\frac{r_S}{\sqrt{E^x_j}}\right) \Delta t_{F,j} \Delta x_j \nonumber\\
               && +\left(
                  \frac{\left(E^x_j\right)'}{4 E^x_j}\frac{1}{1-\frac{r_S}{\sqrt{E^x_j}}} -\left(1-\frac{r_S}{\sqrt{E^x_j}}\right) \left(\frac{\Delta a_j}{\Delta x_j}\right)^2
                  \right)\Delta x^2_j
\end{eqnarray}
and setting,
\begin{equation}
  \Delta a_j =\frac{F_j}{\rho_j} \frac{\left(E^x_j\right)'}{2\sqrt{E^x_j}}\frac{1}{N_j} \frac{1}{1-\frac{r_S}{\sqrt{E^x_j}}}\ell_{\rm Planck},\label{31}
  \end{equation}
one can see that the intervals $\Delta s^F_j$ computed in (\ref{29}, \ref{31}) coincide and therefore we have shown that the line element is invariant. 

This allows to construct $a(x_j)$ from $a(x_S)$ and one recovers the classical change of coordinates when $\ell_{\rm Planck}$ is taken to be infinitesimally small. The polygonal line element is invariant up to all orders in terms of Planck's length. 

This illustrates the invariance when one changes coordinates that imply a change of slicing. It is clear that changes that preserve the foliation keep the line element invariant, provided they are well defined. For example, let us consider the ``tortoise'' coordinate defined as
\begin{eqnarray}
\left(1-\frac{r_S}{x_j}\right)\left(x^*_{j+1}-x^*_j\right)^2 &=& \left(1+\frac{\ell_{\rm Planck}}{2 x_j}\right)^2 
\left(1-\frac{r_S}{x_j}\right)^{-1} \left(x_{j+1}-x_j\right)^2,
\end{eqnarray}
or equivalently,
\begin{equation}
    x^*_{j+1}-x^*_j = \left(1+\frac{\ell_{\rm Planck}}{2 x_j}\right)\left(1-\frac{r_S}{x_j}\right)^{-1}\ell_{\rm Planck},
\end{equation}
where we replaced $x_{j+1}-x_j=\ell_{\rm Planck}$ for all $j$. This equation determines all $x^*_j$ provided the value of, for instance, $x^*_S$. Again, we note that this change of radial coordinate leaves invariant the line element, by construction.

Up to now we have analyzed the case $k_j=(|j|+j_0)^2$ and ignoring superpositions in the quantum states. It is easy to extend the analysis to a general case of $\vert M, \vec{k}\rangle$. If we define $x_j=\sqrt{k_j}\ell_{\rm Planck}$ with a non-uniform spacing, with the discrete interval defined as $(\Delta s_j)^2=g_{ab}(x_j) \Delta x_j^a \Delta x_j^b$ and $\Delta x_j = \sqrt{k_{j+1}}-\sqrt{k_j}$ in the gauge $E^x_j={\rm sig}(j) x^2_j$, the previous proof can be extended easily. 

For a generic superposition states,
\begin{equation}
    \int d M \sum_k c\left(\vec{k}, M\right) \vert M, \vec{k}\rangle,
\end{equation}
the proof can be extended since $\hat{M}, \hat{E}^x$ commute and therefore $\hat{g}_{ab}$ can be defined without ordering ambiguities. One can see that the intervals $\Delta s^S_j$ and $\Delta s^F_j$ (\ref{29},\ref{31}) coincide for any element of the physical basis $|M,\vec k\rangle$ and therefore
their expectation value coincide for any element of the physical space of states. However, due to fluctuations, as it is usual in quantum mechanics, even though the expectation value $\langle{(\widehat{\Delta s_j}})^2\rangle=\langle{\widehat{g_{ab}(t_j,x_j)}}\widehat{\Delta {x^a_j}} \widehat{\Delta {x^b_j}}\rangle$ is invariant, $\langle{\widehat{g_{ab}(t_j,x_j)}}\rangle\langle\widehat{\Delta {x^a_j}}\rangle\langle \widehat{\Delta {x^b_j}}\rangle$ it is not. Thus in a highly quantum regime the length of a curve is gauge invariant but there will be correction to the tensorial behavior of the metric. The previous analysis then provides an explicit and operational notion of quantum covariance. 

Let us now address the covariance of the framework at the bounce that replaces the singularity in \cite{extension}. It is important to remark that the bounce occurs at a point that may be identified in a way that is invariant under changes of foliation and radial coordinates and is given by the infimum of $\vert E_j^x \vert$. Thus, the bounce hypersurface is slicing independent and covariantly defined being the geometric description unique up to this point and, as we shall see, beyond. We start from Schwarzschild's metric given in Eqs. \eqref{eq:schld-gtt}, \eqref{eq:schld-gtx} and \eqref{eq:schld-gxx}. Note that, for this metric, the region $j<r_S/\ell_{\rm Planck}-j_0$ is foliated by $x_j = const$ hypersurfaces, i.e. a nonstationary slicing. However, as we will see, our discussion about covariance is still valid. \footnote{Let us note that it is not difficult to carry out the proof of covariance at the bounce starting from horizon-penetrating coordinates stationary at the exterior.  We adopt the nonstationary Schwarschild's metric for the sake of simplicity.}  At the bounce $x_0=j_0\ell_{\rm Planck}$, we have that,
\begin{eqnarray}
    g^S_{tt}\left(x_0\right)&=& -\left(1-\frac{r_S}{j_0\ell_{\rm Planck}}\right),\\
    g_{tx}^S\left(x_0\right)&=& 0,\\
    g_{xx}^S\left(x_0\right)&=& \left(\frac{\left(E^x_0\right)'}{4E^x_0}\right)^2 \left({1-\frac{r_S}{j_0 \ell_{\rm Planck}}}\right)^{-1}.
\end{eqnarray}

Explicitly, acting on a state, 
\begin{equation}
\left(E^x_0\right)'=\frac{\left(j_0 \ell_{\rm Planck}+\ell_{\rm Planck}\right)^2 -j_0^2 \ell_{\rm Planck}^2}{\ell_{\rm Planck}}=\left(2 j_0+1\right)\ell_{\rm Planck},
\end{equation}
and therefore,
\begin{equation}
    \frac{\left(\left(E^x_0\right)'\right)^2}{4 E^x_0}=\frac{\left(2 j_0+1\right)^2}{4 j_0}=\left(1+\frac{\ell_{\rm Planck}}{2 j_0\ell_{\rm Planck}}\right),
\end{equation}
and 
\begin{equation}
g^S_{xx}\left(x_0\right)=\left(1+\frac{1}{2 j_0}\right)^2 \left(1-\frac{r_S}{j_0 \ell_{\rm Planck}}\right)^{-1}.
\end{equation}

Whereas at $x_{-1}$, the point beyond where the classical singularity would have been (recall (\ref{8})),
\begin{eqnarray}
    g^S_{tt}\left(x_{-1}\right)&=& -\left(1-\frac{r_S}{\left(j_0+1\right)\ell_{\rm Planck}}\right),\\
    g^S_{tx}\left(x_{-1}\right)&=& 0,
\end{eqnarray}
and 
\begin{eqnarray}
    {\left(E^x_{-1}\right)'}&=&\frac{j_0^2\ell_{\rm Planck}^2-\left(j_0+1\right)^2\ell_{\rm Planck}^2}{\ell_{\rm Planck}}=-\left(2 j_0+1\right)\ell_{\rm Planck},\\
    \frac{\left(\left(E^x_{-1}\right)'\right)^2}{4 E^x_{-1}}&=&
    \frac{\left(2 j_0+1\right)^2}{4 \left(j_0+1\right)^2}=
    \frac{4 j_0^2+4 j_0+1}{4\left(j_0+1\right)^2}=
    \left(1-\frac{1}{2\left(j_0+1\right)}\right)^2,
\end{eqnarray}
and as a consequence,
\begin{equation}
   g_{xx}^S\left(x_{-1}\right)=\left(1-\frac{1}{2 \left(j_0+1\right)}\right)^2 \left(1 -\frac{r_S}{\left(j_0+1\right)\ell_{\rm Planck}}\right)^{-1}.
\end{equation}

Let us now consider the generic stationary metric $g^F_{ab}$ with $F(x_j)\in [-1,1]$. We will see that for the system to describe correctly the bounce $F(x_0)$ must be close to one (as usual, not all gauge choices allow to reach the singularity, in this case, the bounce). We start with the general expression of the metric Eqs. \eqref{eq:Fgxx}, \eqref{eq:Fgtt} and \eqref{eq:Fgtx}, and evaluate them at the bounce, namely,
\begin{eqnarray}
    g_{tt}^F\left(x_0\right)&=& 
    -\left(1-\frac{r_S}{j_0\ell_{\rm Planck}}\right),\\
    g_{tx}^F\left(x_0\right)&=&
    \sqrt{\frac{\pi}{\Delta}} {\left(E^x_0\right)'\sqrt{[F(x_0)]^2}}
    \left(1-\frac{r_S}{j_0\ell_{\rm Planck}}+\frac{4\pi j_0^2\ell_{\rm Planck}^2 [F(x_0)]^2}{\Delta}\right)^{-1/2}\nonumber\\
    &=&\sqrt{\frac{\pi}{\Delta}} \left(2j_0+1\right)\ell_{\rm Planck} \sqrt{ [F(x_0)]^2}\left(1-\frac{r_S}{j_0\ell_{\rm Planck}}+\frac{4 \pi j_0^2\ell_{\rm Planck}^2 [F(x_0)]^2}{\Delta}\right)^{-1/2},\\
    g_{xx}^F\left(x_0\right)&=& 
    \left(1+\frac{1}{2 j_0}\right)^2
    \left(1-\frac{r_S}{j_0\ell_{\rm Planck}}+\frac{4 \pi j_0^2\ell_{\rm Planck}^2 [F(x_0)]^2}{\Delta}\right)^{-1},
\end{eqnarray}
with $[F(x_0)]^2 > \left(r_S-j_0\ell_{\rm Planck}\right)\Delta/(4 \pi j_0^3\ell_{\rm Planck}^3)$.

The metric at $x_{-1}$, i.e. $j=-1$ is,
\begin{eqnarray}
    g_{tt}\left(x_{-1}\right) &=& -\left( 1-\frac{r_S}{(j_0+1)\ell_{\rm Planck}}\right),\\
    g_{tx}\left(x_{-1}\right) &=& 
    -\sqrt{\frac{\pi}{\Delta}} \left(2j_0+1\right)\ell_{\rm Planck}\sqrt{[F(x_{-1})]^2}\nonumber\\
    &&\times\left(1 -\frac{r_S}{(j_0+1)\ell_{\rm Planck}}+
    \frac{4 \pi (j_0+1)^2\ell_{\rm Planck}^2 [F(x_{-1})]^2}{\Delta}
    \right)^{-1/2},
\end{eqnarray}
now with $[F(x_{-1})]^2 > \left(r_S-(j_0+1)\ell_{\rm Planck}\right)\Delta/(4 \pi (j_0+1)^3\ell_{\rm Planck}^3)$. Notice that $g_{tx}^F$ changes sign at the bounce since $\left(E^x_0\right)'$ is positive and $\left(E^x_{-1}\right)'$ is negative. This does not introduce singularities in the curvature, as we have shown explicitly in \cite{extension}, where we proved that it is of order Planck at the bounce.

For the spatial component we have that,
\begin{equation}
    g_{xx}^F\left(x_{-1}\right)= 
    \left(1 -\frac{1}{2\left(j_0+1\right)}\right)^2
    \left(1-\frac{r_S}{(j_0+1)\ell_{\rm Planck}}+
    \frac{4 \pi \ell_{\rm Planck}^2 (j_0+1)^2 [F(x_{-1})]^2}{\Delta}\right)^{-1}.
\end{equation}

Making the substitution $t_F(x_0)=t(x_0)-a(x_0)$, the invariant line element is,
\begin{equation}
    \Delta a(x_0) = \frac{F(x_0)}{\rho_0} \frac{\left(E^x_0\right)'}{2 \sqrt{E^x_0}}\left(1 +\frac{[F(x_0)]^2}{\rho_0^2}-\frac{r_S}{\sqrt{E^x_0}}\right)^{-1/2}
    \left(1-\frac{r_S}{\sqrt{E^x_0}}\right)^{-1} \ell_{\rm Planck},
\end{equation}
and $\Delta a(x_{-1})$ is identical substituting $x_0\to x_{-1}$. $\Delta a$ changes sign but is continuous when $\ell_{\rm Planck}$ is taken to be infinitesimally small. 

Following the arguments discussed above, one can easily show that this notion of quantum covariance are immediately applicable for all nodes with $j<0$. The only difference arise in a global sign in \eqref{31}, which indicates that this region is covered by outgoing coordinates if one starts with ingoing coordinates at $j>0$ (and viceversa). 

We have also studied the covariance of several curvature scalars: the Ricci and the Kretschmann scalars, and the scalar obtained by contracting the Weyl tensor with itself. We checked that in the approximation where $x_j$ is treated as a continuous variable, which allows to use derivatives instead of finite differences, these scalars do not depend on the choice of the gauge function $F(x)$. This gives robustness to our model regarding its covariance. It remains to be checked if the discrete version of these scalars is also slicing independent. Nevertheless, the ideas presented in this manuscript regarding the invariance of the spacetime line element of a discrete quantum geometry opens the possibility of studying the covariance of discrete versions of curvature operators and the invariance of curvature scalars.

Given the granularity of space time at the Planck scale, quantum gravity should provide a new principle that replaces general covariance.  But it must still obey certain consistency conditions related to independence of physical effects on the frames we are using, provided these frames are realizable in the quantum theory.  Reference frames
are associated to physical observations, by a system of observers: at rest, free falling or others. In a quantum theory of gravity not all reference frames will  be physically implementable. The holonomization condition that takes an extrinsic curvature of the form $\sin({\rho_j} {K_{\varphi}(x_j)})=F(x_j)$ with $|F(x_j)| \le 1$ provides for each $F$ an explicit definition for the realizable foliations. Also notice that covariance allows to eliminate some ambiguities. For instance in principle it could be possible to choose different polimerizations for the shift and the spatial metric as we did in \cite{us-improv} that would not lead to a quantum covariant formulation. The covariant version of the improved quantization appears in \cite{extension}. 

We have shown here that there exists a quantum operator extension of the line element whose expectation value in any state is independent on the quantum stationary foliation chosen. This provides an explicit and operational notion of quantum covariance that reproduces the usual one at the classical limit. When quantum reference frames are considered, the relation among coordinates associated to two different frames have quantum nature and depend on the observables and c-number functions that describe the change of reference frame. In the explicit case of spherical symmetry in which the radial coordinate is quantized by $x_j$ once the stationary foliation $F(x_j)$ is specified, the description of a given curve whose invariant length we want to evaluate in two different coordinate systems is given in terms of an operator ${\hat t}(O,M,S,j)$ whose form we have determined.   The use of parameterized observables for the coordinate dependent quantities and operatorial change of coordinates as considered in this paper should be present in any approach to quantum gravity. Although we have only shown covariance for the line element for generic stationary slicings, it is likely that it can also be shown for other non-stationary foliations of space-time and scalar quantities that are functions of the geometry. It opens the possibility of discussing covariance in the presence of a discrete geometry. These ideas are not restricted to loop quantum gravity or spherical symmetry.

We wish to thank Martin Bojowald for many useful comments.
This work was supported in part by Grant NSF-PHY-1903799, funds of the
Hearne Institute for Theoretical Physics, CCT-LSU, Pedeciba, Fondo Clemente Estable
FCE 1 2019 1 155865 and the Spanish Government through the projects FIS2017-86497-C2-2-P, PID2019-105943GB-I00 (with FEDER contribution), and the ``Operative Program FEDER2014-2020 Junta de Andaluc\'ia-Consejer\'ia de Econom\'ia y Conocimiento'' under project E-FQM-262-UGR18 by Universidad de Granada.

\end{document}